\documentclass[aps,prc]{revtex4-1}
\usepackage{latexsym}
\usepackage[english]{babel}
\usepackage{graphicx,amsmath,amsfonts,amssymb,amsbsy}
\usepackage{color}
\usepackage{subfigure}
\usepackage{multirow}
\usepackage{mathrsfs}
\usepackage[normalem]{ulem}

\newcommand{\msun}{M_{\odot}}

\newcommand{\ptrans}{p_{\rm trans}}

\newcommand{\etrans}{\varepsilon_{\rm trans}}

\newcommand{\De}{\Delta}
\newcommand{\ep}{\varepsilon}

\newcommand{\cQMsq}{c^2_{\rm QM}}
\newcommand{\Msolar}{M_{\odot}}

\begin{document}

\author{Ignacio F. Ranea-Sandoval}\email{iranea@fcaglp.unlp.edu.ar}
  \affiliation{CONICET, Rivadavia 1917, 1033 Buenos Aires,
  Argentina;\\ Grupo de Gravitaci\'on,
  Astrof\'isica y Cosmolog\'ia,\\ Facultad de Ciencias
  Astron{\'o}micas y Geof{\'i}sicas, Universidad Nacional de La
  Plata,\\ Paseo del Bosque S/N (1900), La Plata, Argentina.}

  \author{Sophia Han} \email{jhan@physics.wustl.edu}
  \affiliation{Department of Physics and Astronomy, University of
    Tennessee, Knoxville, TN~37996, USA}\affiliation{Physics Division,
    Oak Ridge National Laboratory, Oak Ridge, TN~37831, USA}
  \affiliation{Physics Department, Washington University, St.~Louis,
    MO~63130, USA. }

\author{Milva G. Orsaria} \email{morsaria@fcaglp.unlp.edu.ar}
\affiliation{CONICET, Rivadavia 1917, 1033 Buenos Aires,
  Argentina;\\ Grupo de Gravitaci\'on, Astrof\'isica y
  Cosmolog\'ia,\\ Facultad de Ciencias Astron{\'o}micas y
  Geof{\'i}sicas, Universidad Nacional de La Plata,\\ Paseo del Bosque
  S/N (1900), La Plata, Argentina. }

  \author{Gustavo A. Contrera} \email{contrera@fisica.unlp.edu.ar}
\affiliation{CONICET, Rivadavia 1917, 1033 Buenos Aires, Argentina}
\affiliation{IFLP, CONICET - Dpto. de F{\'i}sica, UNLP, La Plata,
  Argentina} \affiliation{Grupo de Gravitaci\'on, Astrof\'isica y
  Cosmolog\'ia,\\ Facultad de Ciencias Astron{\'o}micas y
  Geof{\'i}sicas, Universidad Nacional de La Plata,\\ Paseo del Bosque
  S/N (1900), La Plata, Argentina.}

\author{Fridolin Weber} \email{fweber@mail.sdsu.edu} \affiliation{Department
  of Physics, San Diego State University, 5500 Campanile Drive, San
  Diego, California 92182} \affiliation{Center for Astrophysics and
  Space Sciences, University of California,\\ San Diego, La Jolla, CA
  92093, USA.}

\author{Mark G. Alford} \email{alford@wuphys.wustl.edu}
\affiliation{Physics Department, Washington University,
St.~Louis, MO~63130, USA.}

\title{The Constant-Sound-Speed parameterization for NJL models of
  quark matter in hybrid stars}

\begin{abstract}
The discovery of pulsars as heavy as 2 solar masses
has led astrophysicists to rethink the core compositions of neutron
stars, ruling out many models for the nuclear equations of state
(EoS).  We explore the hybrid stars that occur when hadronic matter is
treated in a relativistic mean-field approximation and quark matter is
modeled by three-flavor local and non-local Nambu$-$Jona-Lasinio (NJL)
models with repulsive vector interactions.  The NJL models typically
yield equations of state that feature a first-order transition to
quark matter. Assuming that the quark-hadron surface tension is high
enough to disfavour mixed phases, and restricting to EoSs that allow
starts to reach 2 solar masses, we find that the appearance of the
quark matter core either destabilizes the star immediately (this is
typical for non-local NJL models) or leads to a very short hybrid star
branch in the mass-radius relation (this is typical for local NJL
models).  Using the Constant-Sound-Speed parametrization we can see
that the reason for the near-absence of hybrid stars is that the
transition pressure is fairly high and the transition is strongly
first-order.
\end{abstract}



\maketitle

\section{Introduction}
\label{sec:intro}

The composition and properties of neutron stars, which are among the
densest objects in the Universe, are not well understood
theoretically. However, measurements of the masses and radii of these
stars are putting strong constraints on the equation of state (EoS),
ruling out many theoretical models proposed over the years or at least
restricting the allowed values of their free parameters.  Specially
the recent discovery of two neutron stars with masses $ \sim 2 \msun$,
by means of the Shapiro Delay for J1614-2230 \cite{Demorest2010} and
white dwarf spectroscopy for J0348+0432 \cite{Antoniadis13}, imposes
severe constraints on models for the EoS of super-dense matter in the
cores of neutron stars
\cite{Bonanno:2011ch,Orsaria:2013hna,Bhowmick:2014,weber:2014,kojo:2015}.

Inside cold neutron stars, the nuclear matter, which is subject to
extreme conditions of high density but low temperature, could undergo
a phase transition, melting hadrons into quarks.  Theoretical
calculations have suggested that the deconfinement transition of
hadronic matter at low temperature but high density is of first order
\cite{Fodor:2004,Casalbuoni:2004}. For neutron stars, this can be
modeled with a hybrid EoS, which describes neutron star matter at low
densities in terms of hadrons and of deconfined quarks at high
densities,
assuming a sharp first-order phase transition between hadronic and
quark matter and considering a generic parametrization of the quark
matter EoS with a density-independent speed of sound, known as the
``Constant-Speed-of-Sound'' (CSS) parametrization
\cite{Haensel:1987,Alford:2013aca}.  We use the CSS parameterization
to gain insight into the range of possible hybrid stars that is
predicted by local and non-local Nambu$-$Jona-Lasinio (NJL) models of
quark matter combined with a relativistic mean field model of nuclear
matter.

Previous analyses of hybrid stars using NJL models of the quark matter
EoS have employed various types of the NJL model, including models
with vector interactions and diquark channels to allow for color
superconductivity \cite{Klahn:2006iw,Bonanno:2011ch,Klahn:2013kga},
models with 8 quark interactions \cite{Benic:2014iaa,Benic:2014jia},
and non-local NJL models
\cite{Blaschke:2007ri,Benic:2013eqa,Orsaria:2013hna,Yasutake:2014oxa}.
Quark matter has also been treated by other methods, such as
perturbative QCD extrapolated down to the densities of interest
\cite{Kurkela:2010yk}, bag models
\cite{Schertler:2000xq,Banik:2002kc,Alford:2004pf,Klahn:2015mfa}, the
Field Correlator Method \cite{Alford:2015dpa}, and Dyson-Schwinger
equations \cite{Klahn:2009mb}.

In this paper we describe quark matter by the widely-studied NJL model
\cite{Vogl:1991qt,Klevansky:1992qe,Hatsuda:1994pi,Buballa:2003qv} with
both local and non-local interactions. The non-local NJL model
includes a form factor in the quark-quark interactions that can be
used to simulate the effects of confinement (by generating a quark
propagator without poles at real energies
\cite{Praschifka:1989fd,Ball:1990wq,Schmidt:1994di,Plant:1997jr}) and
also functions as a natural regulator.  In fact, non-locality arises
naturally in the context of several successful approaches to
low-energy quark dynamics as, for example, one gluon exchange
\cite{Ripka:1997zb,GomezDumm:2006vz,Contrera:2007wu} (as in this
work), the instanton liquid
model~\cite{Diakonov:1985eg,Schafer:1996wv} and the Schwinger-Dyson
resummation techniques~\cite{Roberts:1994dr,Roberts:2000aa}.  Lattice
QCD calculations~\cite{Parappilly:2005ei} also indicate that quark
interactions should act over a certain range in momentum
space. Moreover, several
studies~\cite{Bowler:1994ir,Plant:1997jr,Broniowski:2001cx,Rezaeian:2004nf,Contrera:2009hk}
have shown that non-local chiral quark models provide a satisfactory
description of hadron properties at zero temperature and density.  In
our NJL calculations we also include vector interactions and the
formation of a chiral condensate, but not diquark condensation (color
superconductivity \cite{Alford:2007xm}).

The main questions which are being addressed in this paper are as
follows.  Does the range of physically plausible parameters of the NJL
quark matter EoS allow for hybrid stars, and if so how long is the
hybrid star branch, and are there observable signatures such as a
disconnected third-family branch of hybrid stars as noted in
Refs.~\cite{SchaffnerBielich:2002ki,Agrawal:2010er,Benic:2014jia}?
Does the CSS parameterization adequately capture the functional form
of the NJL quark matter EoS? Can the CSS parameterization be used to
understand the mass-radius characteristics of NJL hybrid stars?

In particular, previous analysis of hybrid stars using the CSS
parameterization of possible quark matter EoSs has shown that it is
easiest to achieve masses of $2\,\Msolar$ and above if the quark
matter is stiff, with a high speed of sound $\cQMsq \gtrsim 0.4$
\cite{Alford:2015dpa}.  If the hadron-quark transition is at low
pressure then a long hybrid branch can exist in the mass-radius plane;
if the transition is at high pressure then the hybrid branch, if it
exists, tends to be very small. In this paper we will see that the NJL
models studied here are of the ``high transition pressure'' type.

The paper is organized as follows. In section \ref{sec:dense} we
describe briefly the models for the hadron and quark phases of the
hybrid stars.  In section \ref{sec:cssparam} we review the CSS
parametrization proposed in \cite{Alford:2013aca} and check to what
degree it can accurately characterize our NJL EoSs.  In section
\ref{sec:mass-radius} we discuss the hybrid stars resulting from our
NJL models, and in section \ref{sec:end} we give our conclusions.

\section{Dense matter in the cores of hybrid stars}
\label{sec:dense}

\subsection{The outer core}

We model the matter in the outer core of a hybrid star using a
non-linear relativistic mean field theory
\cite{Walecka:1974,boguta77:a,boguta77:b,boguta83:a,Serot:1986}, where
baryons (neutrons, protons, hyperons and delta states) interact via
the exchange of scalar, vector and isovector mesons ($\sigma $,
$\omega$ and $\rho $ mesons, respectively). Replacing all baryon
currents in the field equations with their respective ground-state
expectation values, we arrive at a system of mean-field equations at
zero temperature given by
\begin{eqnarray}
m_{\sigma}^2 \sigma&=& -\frac{dU}{d \sigma} + \sum_{B}
\frac{2J_B+1}{2\pi^2} g_{\sigma B} \int_{0}^{k_B} p^2 dp
\frac{m_B-g_{\sigma B} \sigma}{[p^2+(m_B-g_{\sigma
      B}\sigma)^2]^{1/2}}\,\,\, , \nonumber\\ m_{\omega}^2 \omega_0&=&
\sum_{B} g_{\omega B} n_{B}\,\,\, , \nonumber
\\ m_{\rho}^2\rho_{03}&=& \sum_{B}g_{\rho B}I_{3B} n_{B}\,\,\,
, \label{eq:hadron}
\end{eqnarray}
where $I_{3B}$ and $J_B$ are the 3-components of isospin and
    spin, respectively, and $k_B$ is the Fermi momentum of a baryon of
    type $B$. The baryon-meson coupling constants $g_{\sigma
          B}=x_{\sigma B}g_{\sigma}$, $g_{\omega B}=x_{\omega
          B}g_{\omega}$, and $g_{\rho B}=x_{\rho B}g_{\rho}$ are
        expressed in terms of the scalar, vector, and isovector
        coupling constants $g_{\sigma}$, $g_\omega$, and
        $g_\rho$ of the hadronic model parametrizations (GM1 and NL3) studied in this paper \cite{Glendenning:1991es,Lalazissis:1996rd,Orsaria:2013hna}.
    Following Ref.\ \cite{Glendenning:1985} we take $x_{\sigma
      B}=0{.}7$ and $x_{\omega B}=x_{\rho B}=1$. The quantity $U$ in Eq.\ (\ref{eq:hadron})
    denotes the non-linear $\sigma$-meson potential, representing the
    self-interactions of the scalar $\sigma$ field. It can be written
    as
\begin{equation}
U=[b_1 \, m_N + b_2 \, (g_{\sigma} \sigma)] \, (g_{\sigma} \sigma)^3,
\label{eq:potu}
\end{equation}
with $b_{1,2}$ denoting constants determined by the properties of
hadronic matter. The quantity $m_N$ is the nucleon mass. Solving
Eq.\ (\ref{eq:hadron}) together with the charge neutrality condition
and baryonic number conservation, we obtain the total pressure and the
energy density of the hadronic matter in the outer core of the hybrid
star (for details, see Refs.\ \cite{Glendenning:1985,weber1999:a}).

\subsection{The inner core}

To describe the matter in the inner core of the hybrid star we use a
local and non-local extension of the NJL model. The non-local NJL
extension includes non-local interactions and has several advantages
over the local NJL model (for details see \cite{Orsaria:2013hna} and
references therein).

\subsubsection{The non-local NJL model}

For this model, the mean-field thermodynamic potential at zero
temperature is given by \cite{Orsaria:2013hna}:
\begin{eqnarray}
&&\Omega^{NL} (M_f,0,\mu_f) = -\frac{N_c}{\pi^3}\sum_{f=u,d,s}
  \int^{\infty}_{0} dp_0 \int^{\infty}_{0}\, dp \, p^2 \,\mbox{ ln }\left\{
  \left[\widehat \omega_f^2 + M_{f}^2(\omega_f^2)\right]
  \frac{1}{\omega_f^2 + m_{f}^2}\right\} \nonumber \\ & & -
  \frac{N_c}{\pi^2} \sum_{f=u,d,s} \int^{\sqrt{\mu_f^2-m_{f}^2}}_{0}
  dp\,\, p^2\,\, \left[(\mu_f-E_f) \theta(\mu_f-m_f) \right] \\ & & -
  \frac{1}{2}\left[\sum_{f=u,d,s} (\bar \sigma_f \ \bar S_f +
    \frac{G_S}{2} \ \bar S_f^2) + \frac{H}{2} \bar S_u\ \bar S_d\ \bar
    S_s\right]
- \sum_{f=u,d,s}\frac{\varpi_f^2}{4 G_V} \, , \nonumber
\label{omzerot}
\end{eqnarray}
where $N_c=3$ (as we are considering three quark colors), $E_{f} =
\sqrt{p^{2} + m_{f}^{2}}$, and $\omega_f^2 =
(\,p_0\,+\,i\,\mu_f\,)^2\,+\,p^2$.  The constituent quark masses,
$M_{f}$, are treated as momentum-dependent quantities and are given by
\begin{equation}
M_{f}(\omega_{f}^2) \ = \ m_f + \bar\sigma_f g(\omega_{f}^2)\, ,
\end{equation}
where $g(\omega^2_f)$ denotes a form factor which we take to be
Gaussian, $g(\omega^2_f) = \exp{\left(-\omega^2_f/\Lambda^2\right)}$.

The inclusion of vector interactions shifts the quark chemical
potential by
\begin{equation}
\mu_f\, \rightarrow \widehat{\mu}_f=\mu_f - g(\omega^2_f)\varpi_f\, ,
\label{eq:mu_f}
\end{equation}
where $\varpi_f$ represents the vector mean fields related to the
vector current interaction. The inclusion of the form factor in
Eq.\ (\ref{eq:mu_f}) is a particular feature of the non-local model,
which renders the shifted chemical potential momentum dependent.  The
squared four momenta, $\omega_f^2$, in the dressed part of the
thermodynamic potential are modified as
\begin{equation}
\omega_f^2\,\rightarrow \widehat{\omega}_f^2 = (\,p_0\, +
\,i\,\widehat{\mu}_f\,)^2\, + \,p^2 \, .
\end{equation}

Note that in order to avoid a recursive problem, as discussed in
Refs.\ \cite{Dumm,Weise2011,Contrera:2012wj}, the quark chemical
potential shift does not affect the non-local form factor
$g(\omega_{f}^2)$. In this work we use the
parameters listed in Table\ \ref{tabla:n3NJL_param} for the non-local NJL model.

As was proven in Ref.\ \cite{Scarpettini}, within the stationary phase
approximation, the mean-field values of the auxiliary fields, $\bar
S_f$, are related to the mean-field values of the scalar fields $\bar
\sigma_f$. They are given by
\begin{equation}
\bar S_f = -\, 16\,N_c\, \int^{\infty}_{0}\,dp_0 \int^{\infty}_{0}
dp \, \frac{p^2}{(2\pi)^3} \, g(\omega_f^2)\, \frac{
  M_{f}(\omega_f^2)}{\widehat{\omega}^2 + M_{f}^2(\omega_f^2)}\, .
\end{equation}

The mean field values of $\bar \sigma_f$ and $\varpi_{f}$ are obtained
via minimizing the thermodynamic potential with respect to variations
in these quantities, that is,
\begin{equation}
\label{nonloq}
\frac{\partial \Omega^{NL}}{\partial \bar \sigma_f} = 0\, ,
\quad\frac{\partial \Omega^{\rm NL}}{\partial {\varpi_{f}}}= 0\, .
\end{equation}
The relevant mean-field flavor fields are $\bar \sigma_u$, $\bar
\sigma_d$ and $\bar \sigma_s$.

\begin{table}[htb]
\begin{center}
\begin{tabular}{cccc}
\hline \hline
 Parameters~~~~~~~~ & Set I  & Set II  & Set III\\
 \hline
$\bar m$[MeV]   & $5.0$ &$5.5$ &$6.2$ \\
$m_s$   [MeV]    & $119.3$ &$127.8$ &$140.7$  \\
$\Lambda$ [MeV]  & $843.0$ &$780.6$ &$706.0$ \\
$G_S \Lambda^2$   & $13.34$ &$14.48$ &$15.04$ \\
$H \Lambda^5$     & $-273.75$ &$-267.24$ &$-337.71$ \\
\hline \hline
\end{tabular}
\end{center}
\caption{Sets of parameters used for the non-local NJL model
  calculations presented in this paper.}
\label{tabla:n3NJL_param}
\end{table}

\subsubsection{The local NJL model}

For the local SU(3) NJL model, we use the scheme and parameters of
Refs.~\cite{Orsaria:2012je,Orsaria:2013hna} (and references therein).
At the mean-field level, the thermodynamic potential at zero
temperature reads
\begin{widetext}
\begin{eqnarray}
\Omega^{L}(M_f, \mu) &=& G_S
\sum_{f=u,d,s}\left\langle\bar{\psi_{f}}\psi_{f}\right\rangle^2 + 4
H\left\langle\bar{\psi_{u}}\psi_{u}\right\rangle
\left\langle\bar{\psi_{d}}\psi_{d}\right\rangle\left\langle
\bar{\psi_{s}}\psi_{s}\right\rangle - 2
N_c\sum_{f=u,d,s}\int_\Lambda\frac{\mathrm{d}^3p}
{\left(2\pi\right)^3}\,{E_f} \nonumber \\ && -
\frac{N_c}{3\pi^2}\sum_{f=u,d,s}\int
_0^{p_{F_f}}\,\mathrm{d}{p}\,~\frac{p^4}{E_f} - G_V\,\sum_f \rho_f^2
\, , \label{omega_njl}
\end{eqnarray}
\end{widetext}
where $N_c = 3$, $E_{f} = \sqrt{p^{2}+M_{f}^{2}}$, and
$p_{F_f} = \sqrt{\mu_f^2-M_{f}^{2}}$.  The constituent quark masses
$M_{f}$ are given by
\begin{equation}
M_{f}=m_{f}-2G_S\left\langle\bar{\psi_{f}}\psi_{f}\right\rangle - 2H
\left\langle\bar{\psi_{j}} \psi_{j} \right\rangle\left\langle\bar
                {\psi_{k}} \psi_{k} \right\rangle \, ,
\end{equation}
with $f,j,k=u,d,s$ indicating cyclic permutations.  The vector
interaction shifts the quark chemical potential according to
\begin{equation}
\mu_f\, \rightarrow \mu_f - 2G_V\rho_f \, ,
\end{equation}
where $\rho_f$ is the quark number density of the quark field of
flavor $f$ in the mean field approximation, that is,
\begin{equation}
\rho_f=\frac{N_c}{3\pi}[(\mu_f - 2G_V\rho_f)^2-M_f^2]^{3/2} \, .
\end{equation}
The mean field values are determined from the solution of the gap
equations, obtained by minimizing the thermodynamic potential with
respect to the quark condensates
$\left\langle\bar{\psi_{f}}\psi_{f}\right\rangle$,
\begin{equation}
\label{loq}
\frac{\partial \Omega^{L}}{\partial
  \left\langle\bar{\psi_{f}}\psi_{f}\right\rangle}= 0\, , \quad
f=u,d,s \, .
\end{equation}

\begin{table}[htb]
\begin{center}
\begin{tabular}{ccc}
\hline \hline
 Parameters~~~~~~~~  & Set IV & Set V \\
 \hline
$\bar m$ [MeV]  & $5.5$&$5.5$ \\
$m_s$  [MeV]     & $135.7$& $140.7$ \\
$\Lambda$  [MeV] & $631.4$& $602.3$ \\
$G_S \Lambda^2$   & $3.67$& $3.67$ \\
$H \Lambda^5$     & $-9.29$& $-12.3$ \\
\hline \hline
\end{tabular}
\end{center}
\caption{Sets of parameters used for the local NJL model calculations
  presented in this paper.}
\label{tabla:3NJL_param}
\end{table}

\subsection{The hadron-quark phase transition}
\label{sec:trans}

Several theoretical works \cite{Schertler:2000xq,Fraga:2001xc,%
  Banik:2002kc,Buballa:2003et,Benic:2014jia,Agrawal:2010er,%
  Kurkela:2010yk, Negreiros:2012} have suggested that there might be a
first-order phase transition between hadronic and quark matter at low
temperatures. The density at which such a phase transition occurs is
not known, but it is expected to occur at several times nuclear
saturation density. In a neutron star, such a transition can lead to
two possible structures, depending on the surface tension between
hadronic and quark matter
\cite{Alford:2001zr,lida:2008,endo:2011,Pinto:2012aq,lugones:2013,ke:2014}.

If the surface tension between hadronic and quark matter is bigger
than a critical value, which is estimated to be between around $5$ to
$40$ MeV/fm${}^2$ \cite{Alford:2001zr,endo:2011}, then there is a
sharp interface (Maxwell construction) between neutral hadronic matter
and neutral quark matter. If the surface tension is below the critical
value then there is a mixed phase (Gibbs construction), with
charge-separated domains of quark and hadronic matter occurring over a
finite range of pressures. In this work we will assume that the
surface tension of the interface is high enough to ensure that the
transition occurs at a sharp interface. This is a possible scenario,
given the uncertainties in the value of the surface tension
\cite{Alford:2001zr,Palhares:2010be,Pinto:2012aq}.  For a discussion
of generic equations of state that continuously interpolate between
the phases to model mixing or percolation, see
Refs.~\cite{kojo:2015,Macher:2004vw,Masuda:2012ed}.

\begin{table}[h]
\begin{tabular}{|c|c|c|c|c|c|c|c|c|c|}
\hline
\multicolumn{2}{|c|}{EoS}                       & $G_V/G_S$ & $p_{\rm trans}/\epsilon_{\rm trans}$ & $\Delta
\epsilon/\epsilon_{\rm trans}$& $c_{\rm QM}^2$ & $M_{\rm max}~ [M_\odot]$ & $R_{\rm M}$~ [km] & $\De M ~[M_\odot]~ $(CSS)& $\De M
[M_\odot]~ $(NJL)\\ \hline
\multirow{9}{*}{GM1} & \multirow{3}{*}{Set I}    & 0.00      & \multicolumn{7}{c|}{No Phase Transition}                       \\
\cline{3-10}
                     &                          & 0.05              & 0.25     & 0.78      & 0.13       & 2.10
                     & 13.36    & $5.1 \times 10^{-6}$     &$< 10^{-5}$ \\ \cline{3-10}
                     &                          & 0.09              & 0.27     & 0.83     & 0.26       & 2.17                 &
                     13.19    & $1.4 \times 10^{-6}$  &$< 10^{-5}$   \\ \cline{2-10}
                     & \multirow{3}{*}{Set II}  & 0.00    & 0.24      & 1.21     & 0.22      &2.08                  &13.39     &0
                     &0   \\ \cline{3-10}
                     &                          & 0.05              & 0.28      & 1.09      & 0.27       & 2.18
                     &13.18      &0 &0 \\ \cline{3-10}
                     &                          & 0.09              & 0.31      & 0.96     &0.29      & 2.25              & 12.97
                     &$< 10^{-6}$  &$< 10^{-5}$ \\ \cline{2-10}
                     & \multirow{3}{*}{Set III} & 0.00    & 0.24      & 1.35     & 0.26      & 2.09                & 13.38    &0
                     &0  \\ \cline{3-10}
                     &                          & 0.05              &0.29       & 1.12     & 0.32      & 2.21                  &
                     13.10     &0  &0  \\ \cline{3-10}
                     &                          & 0.09              & 0.33      & 0.87     & 0.46      & 2.29                  &
                     12.78     & $1.1 \times 10^{-6}$ &$< 10^{-5}$ \\ \hline
\multirow{9}{*}{NL3} & \multirow{3}{*}{Set I}   & 0.00      & \multicolumn{7}{c|}{\multirow{3}{*}{No Phase Transition}}      \\
\cline{3-3}
                     &                          & 0.05      & \multicolumn{7}{c|}{}                          \\ \cline{3-3}
                     &                          & 0.09      & \multicolumn{7}{c|}{}                          \\ \cline{2-10}
                     & \multirow{3}{*}{Set II}  & 0.00      & \multicolumn{7}{c|}{No Phase Transition}                       \\
                     \cline{3-10}
                     &                          & 0.05                & 0.27      & 1.46    & 0.17    & 2.37
                     &14.59  &0  &0 \\ \cline{3-10}
                     &                          & 0.09                &0.29      & 1.39    & 0.25     & 2.46
                     &14.49    &0  &0  \\ \cline{2-10}
                     & \multirow{3}{*}{Set III} & 0.00      & 0.24      & 1.87    & 0.23    & 2.27               & 14.66   &0
                     &0   \\ \cline{3-10}
                     &                          & 0.05                & 0.28       & 1.65   & 0.29     & 2.42                &
                     14.54   &0   & 0 \\ \cline{3-10}
                     &                          & 0.09               & 0.32       & 1.42   & 0.36      & 2.54              &14.40
                     &0    &0 \\ \hline
\end{tabular}

\caption{ Properties of the compact stars arising from quark matter
  obeying the non-local NJL EoS. We show results for two hadronic
  EoSs, GM1 and NL3, and for various NJL parameter values (sets I,
  II, III, see Table~\ref{tabla:n3NJL_param}), and vector couplings
  $G_V$. For each case we show the three CSS parameters that
  characterize the quark matter EoS (see Sec.~\ref{sec:cssparam}),
  along with the mass and radius of the heaviest star, and the mass
  range $\De M$ of the hybrid branch, obtained from the CSS parameters
  and from the original non-local NJL EoS.}
\label{table:nolocal}
\end{table}

\begin{table}[h]
\begin{tabular}{|c|c|c|c|c|c|c|c|c|c|}
\hline
\multicolumn{2}{|c|}{EoS}   & $G_V/G_S$ & $p_{\rm trans}/\epsilon_{\rm trans}$ & $\Delta
\epsilon/\epsilon_{\rm trans}$& $c_{\rm QM}^2$ & $M_{\rm max}~ [M_\odot]$ & $R_{\rm M}$~ [km]
& $\De M [M_\odot]~ $(CSS)& $\De M~ [M_\odot]$~ (NJL)\\ \hline
\multirow{6}{*}{GM1} & \multirow{3}{*}{Set IV} & 0.00          & 0.20  & 0.39  & 0.33       & 1.93                  & 13.44
&$5 \times 10^{-2}$      &$1.1 \times 10^{-2}$      \\ \cline{3-10}
                     &                         & 0.15                                  & 0.27  & 0.48  & 0.20       & 2.16
                     & 13.21       & $3 \times 10^{-3}$     &$1.4 \times 10^{-3}$    \\ \cline{3-10}
                     &                         & 0.30                                   & 0.32  & 0.66 & 0.23     & 2.27
                     & 12.90       &$2.5 \times 10^{-4}$      &$2.2 \times 10^{-4}$    \\ \cline{2-10}
                     & \multirow{3}{*}{Set V}  & 0.00                       & 0.25  & 0.60  & 0.22      & 2.10
                     & 13.35       &$7.6 \times 10^{-4}$      &$4.9 \times 10^{-4}$   \\ \cline{3-10}
                     &                         & 0.15                                   & 0.30  & 0.85  & 0.19      &2.23
                     & 13.04       &$1.7 \times 10^{-6}$    &$< 10^{-5}$  \\ \cline{3-10}
                     &                         & 0.30                                   & 0.34  & 0.84  & 0.27    & 2.30
                     & 12.75       &$1.9 \times 10^{-6}$      &$< 10^{-5}$    \\ \hline
\multirow{6}{*}{NL3} & \multirow{3}{*}{Set IV}    & 0.00         & 0.17  & 0.64  & 0.31    & 1.84                  & 14.79
&$2.8 \times 10^{-4}$     &$5.3 \times 10^{-4}$   \\ \cline{3-10}
                     &                         & 0.15                                   & 0.22  & 0.57  & 0.35     & 2.15
                     & 14.70      & $1.8 \times 10^{-3}$     &$2.6 \times 10^{-3}$       \\ \cline{3-10}
                     &                         & 0.30                                   & 0.27  & 0.59  & 0.24     & 2.39
                     & 14.57      &$8.4 \times 10^{-4}$      &$1.1 \times 10^{-3}$      \\ \cline{2-10}
                     & \multirow{3}{*}{Set V}  & 0.00                       & 0.22  & 0.84  & 0.33     & 2.15
                     &14.72        &0    &0        \\ \cline{3-10}
                     &                         & 0.15                                   & 0.27  & 0.84  & 0.21     & 2.39
                     & 14.58        &$2.8 \times 10^{-6}$    &$< 10^{-5}$   \\ \cline{3-10}
                     &                         & 0.30                                   & 0.31  & 0.99  & 0.21     & 2.52
                     & 14.42        &0    &0  \\ \hline
\end{tabular}
\caption{ Properties of the compact stars arising from quark matter
  obeying the local NJL EoS. We show results for two hadronic EoSs,
  GM1 and NL3, and for various NJL parameter values (sets IV, V
  see Table~\ref{tabla:3NJL_param}), and vector couplings $G_V$. For
  each case we show the three CSS parameters that characterize the
  quark matter EoS (see Sec.~\ref{sec:cssparam}), along with the mass
  and radius of the heaviest star, and the mass range $\De M$ of the
  hybrid branch, obtained from the CSS parameters and from the
  original local NJL EoS.  }
\label{table:local}
\end{table}

\section{The CSS parametrization}
\label{sec:cssparam}

\begin{figure}[htb]
\centering
\includegraphics[width=0.8 \textwidth]{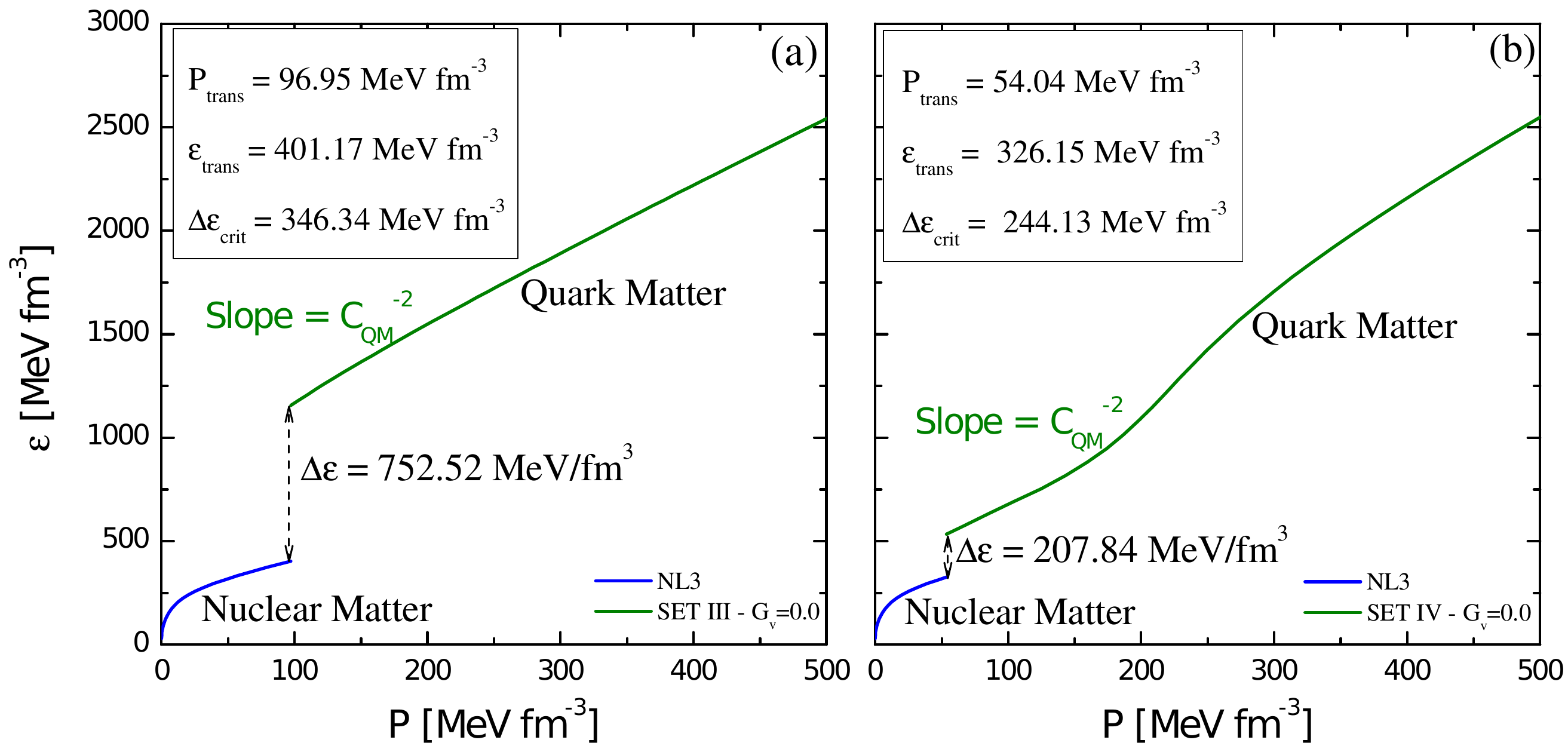}
 \caption{(Color online) EoS, $\ep (p)$, for dense matter without
vector interaction, showing the  energy density discontinuity $\De \ep$
between hadronic matter (NL3) and quark matter.
For quark matter modeled by non-local NJL SET III, panel (a),
the discontinuity is greater than $\De \ep_{\rm crit}$ (\ref{eq:stabalf}),
so there is no stable connected hybrid branch.
In contrast, the local NJL SET IV, panel (b), has a smaller discontinuity
$\De \ep<\De \ep_{\rm crit}$ which leads to a connected but
very short hybrid branch (see Table~\ref{table:local}).}
\label{fig:gapgv0}
\end{figure}

The CSS parameterization assumes that there is a sharp interface
between nuclear matter and quark matter, and that the speed of sound
in quark matter is pressure-independent for pressures ranging from the
first-order transition pressure up to the maximum central pressure of
a neutron star.  The main features of the quark matter EoS can then be
captured by three parameters (see Fig.~\ref{fig:gapgv0}): the pressure
$\ptrans$ at the transition, the discontinuity in energy density
$\De\ep$ at the transition, and the speed of sound $c_{\rm QM}$ in the
high-density phase.  For a given nuclear matter EoS $\ep_{\rm NM}(p)$,
the CSS parameterization of the EoS takes the form
\begin{equation}
\ep(p) = \left\{\!
\begin{array}{ll}
\ep_{\rm NM}(p)\, , & p<p_{\rm trans} \\ \ep_{\rm NM}(p_{\rm trans})+\De
\ep+c_{\rm QM}^{-2} (p-p_{\rm trans}) \, , & p>p_{\rm trans}
\end{array}
\right.\ .
\label{eqn:EoSqmparam}
\end{equation}
The values of the CSS parameters
$(\ptrans/\etrans,\,\De\ep/\etrans,\,\cQMsq)$ for the EoSs described
in Sec.~\ref{sec:dense} are given in Tables \ref{table:nolocal}
(non-local NJL) and \ref{table:local} (local NJL).

The discontinuity in the energy density $\De \ep$ is a measure of how
strongly the first-order phase transition is, and determines the
presence or absence of a connected hybrid star branch.  When the
central pressure rises just above $\ptrans$ one would expect a very
small core of quark matter to appear. However, it can be shown
\cite{Seidov:1971,Haensel:1983,Lindblom:1998} that, {\em independent}
of the speed of sound in quark matter,
\begin{equation}
\begin{array}{rclcl}
\De\ep &>& \De \ep_{\rm crit} &\Rightarrow& \mbox{no connected
  branch,} \\ \De\ep &<& \De \ep_{\rm crit} &\Rightarrow&
\mbox{connected branch,} \\[2ex] \multicolumn{3}{l}{\displaystyle
  \mbox{where}\quad \frac{\De \ep_{\rm crit}}{\etrans}} &=&
\displaystyle \frac{1}{2} + \frac{3}{2} \frac{\ptrans}{\etrans} \ . \\
\end{array}
\label{eq:stabalf}
\end{equation}
In other words, if the energy density jump at the transition is bigger
than $\De \ep_{\rm crit}$ then the quark matter core, no matter how
small, destabilizes the star, and the mass-radius relationship
contains no branch of hybrid stars connected to the hadronic branch.
It must be emphasized strongly that this conclusion is independent of
the speed of sound in quark matter, and that the criterion
(\ref{eq:stabalf}) is valid independent of how pressure-independent
the speed of sound in quark matter is.  It is only for predictions of
the presence or absence of {\em disconnected} branches or for
properties of stable hybrid stars (mass, radius) that the CSS
parameterization relies on the assumption of a pressure-independent
speed of sound.

\begin{figure}[htb]
\centering
\includegraphics[width=0.45 \textwidth]{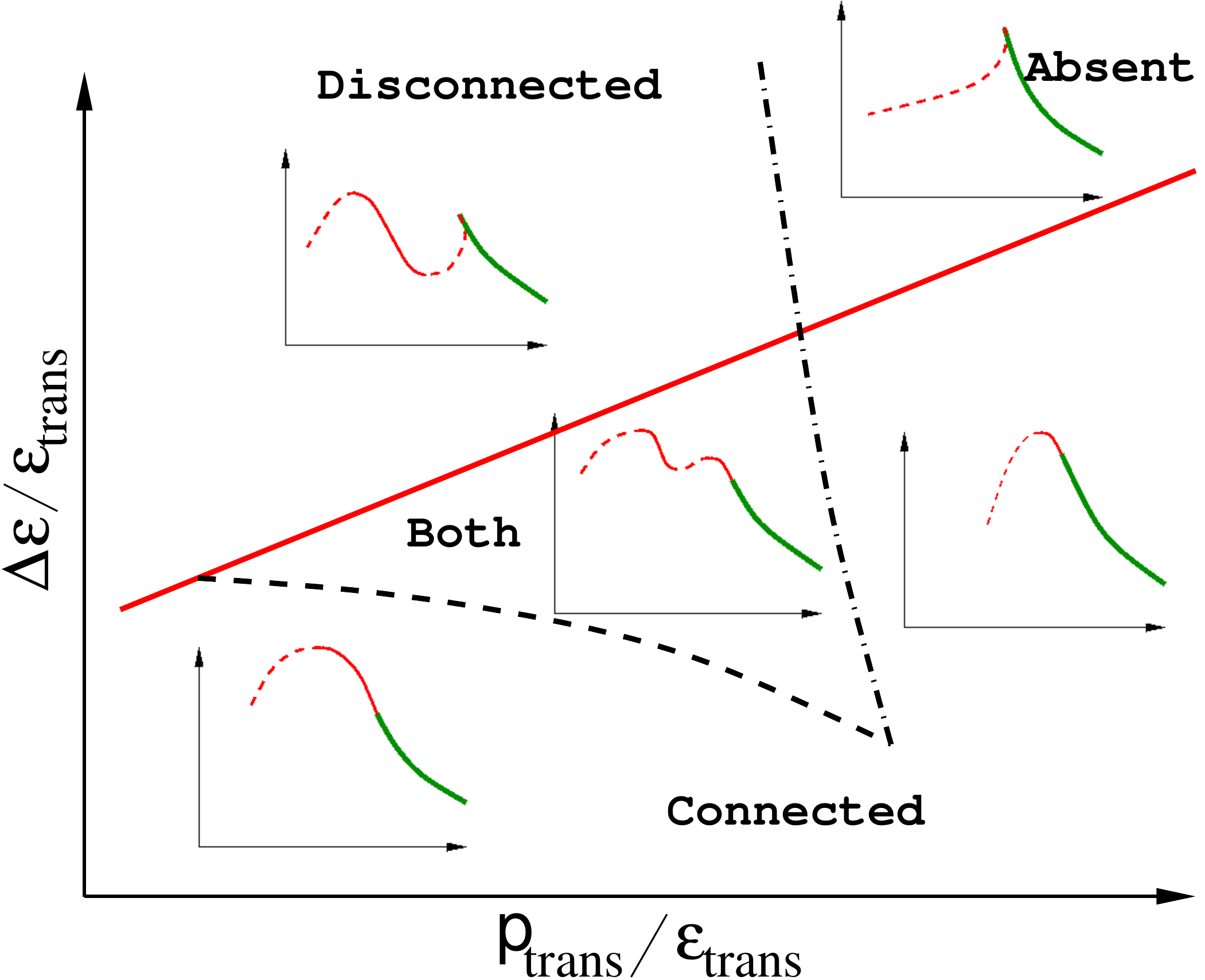}
\caption{(Color online) Schematic phase diagram of the possible
  topologies of M(R) \cite{Alford:2013aca}. The thick straight (red) line
  (Eq.~(\ref{eq:stabalf})) separates EoSs with a connected hybrid
  branch (regions B and C below the line) from those without a
  connected hybrid branch (regions A and D above the line).  The
  dashed black lines delimit the regions (B and D) where disconnected
  hybrid branches occur.  The insets show $M(R)$ in each region, with
  a thin dashed (red) line indicating unstable hybrid stars and a thin
  solid (red) line indicating stable hybrid stars.
\label{fig:schem}
}
\end{figure}



\begin{figure}[htb]
\centering
\includegraphics[width=1.0 \textwidth]{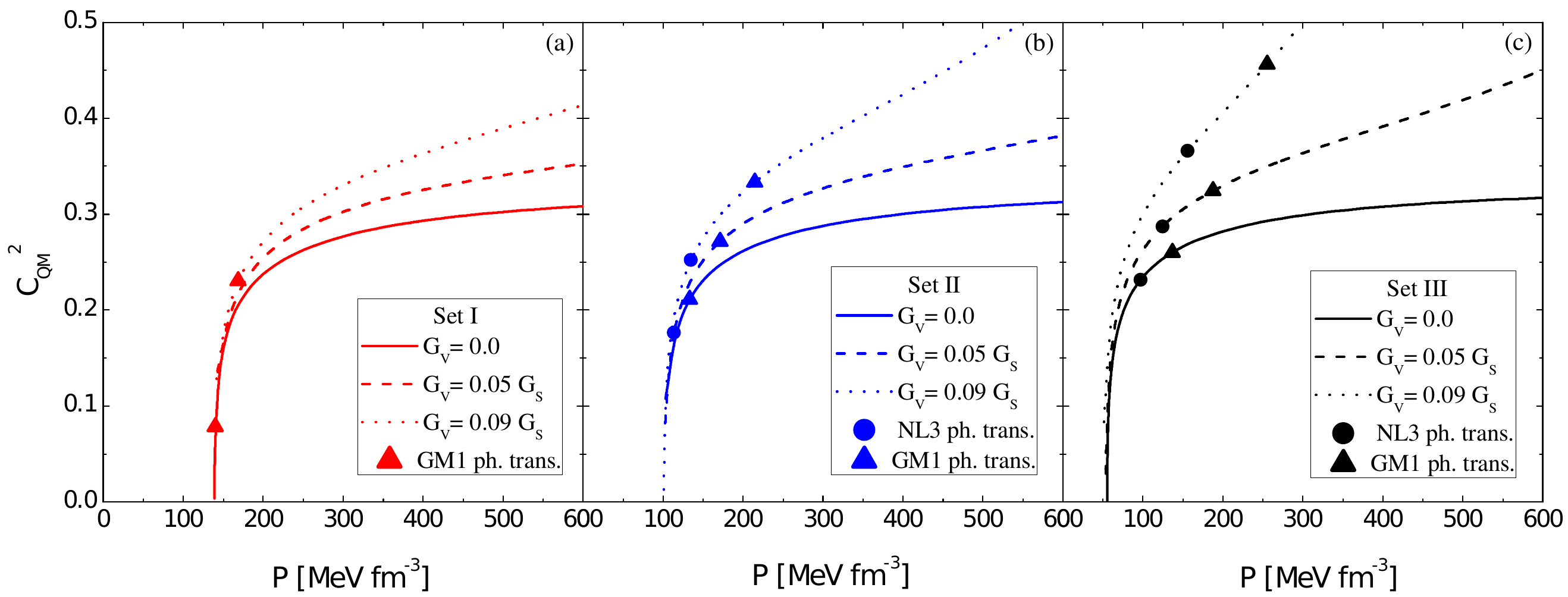}
\caption{ (Color online) Squared speed of sound for the non-local NJL
  quark matter model as a function of pressure.  NL3 to quark matter
  transitions are marked by dots; GM1 $\to$ quark matter is marked by
  triangles.  Solid lines show the behaviour without vector
  interactions, while dashed and dotted lines represent the results
  for a finite vector interaction coupling constant. Left, central and
  right panels correspond to the non-local NJL model with parameter
  set I, panel (a), set II, panel (b), and set III, panel (c).}
\label{fig:Cs2_nonlocal}
\end{figure}

\begin{figure}[htb]
\centering
\includegraphics[width=0.65 \textwidth]{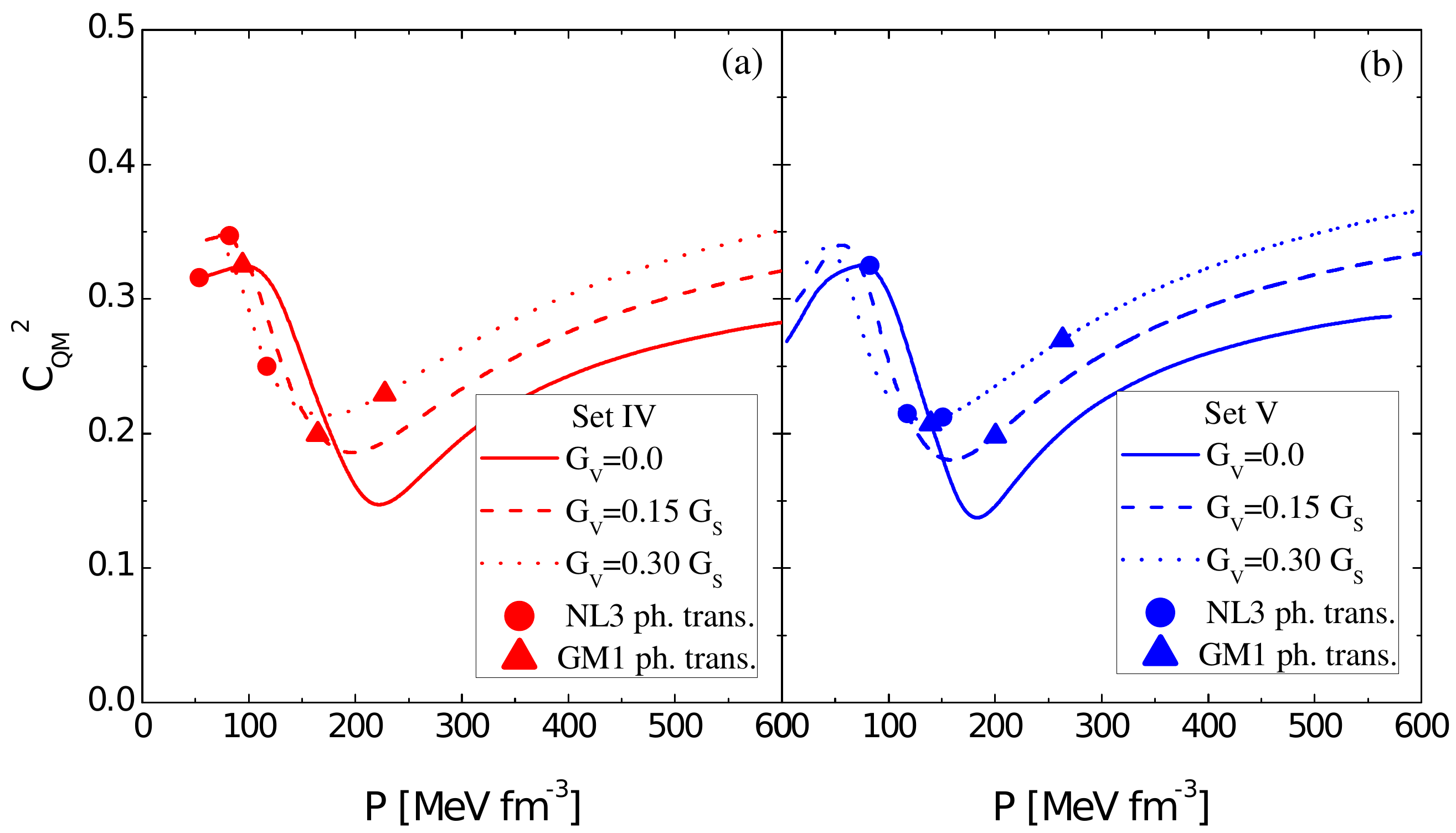}
\caption{ (Color online) Squared speed of sound for the local NJL
  quark matter model as function of pressure.  The symbols and line
  styles are similar to those in Fig.~\ref{fig:Cs2_nonlocal}. Left and
  right panels correspond to the local NJL model with parameter set
  IV, panel (a), and set IV, panel (b). Note that the local NJL model shows a more
  density-independent speed of sound, and thus is more accurately
  characterized by the CSS parameterization than in the non-local
  case.}
\label{fig:Cs2_local}
\end{figure}

In Figs.~\ref{fig:Cs2_nonlocal} and \ref{fig:Cs2_local} we plot
$\cQMsq$ as a function of the pressure for all the quark matter EoSs
considered in Sec.~\ref{sec:dense}. We also show where the hadron to
quark phase transitions occur, as dots (for the NL3 hadronic EoS) and
triangles (for the GM1 hadronic EoS). NL3 is a stiffer EoS, and it
undergoes the phase transition at lower $\ptrans$.

In the local NJL models (Fig.~\ref{fig:Cs2_local}) the speed of sound
is only mildly sensitive to the NJL model parameters such as $m_s$
(which increases from Set IV to Set V) and the vector coupling
constant $G_V$.  Moreover, above the hadron to quark phase transition
the speed of sound shows only a moderate dependence on pressure, tending
to stay within the range $\cQMsq \sim 0.2$ to 0.3. We therefore expect
that the local NJL models will be reasonably accurately characterized
by the CSS parameterization.

The non-local NJL model (Fig.~\ref{fig:Cs2_nonlocal}), in contrast,
shows great sensitivity to the NJL model parameters. As $m_s$ and
$G_V$ increase we see that $\cQMsq$ varies more strongly with
pressure. For higher values of $m_s$ and $G_V$ the non-local NJL model
EoSs are outside the family of EoSs that can be accurately
characterized by the CSS parameterization.  However, as noted above,
the criterion (\ref{eq:stabalf}) for the presence of connected hybrid
branches is independent of the speed of sound; the density dependence
of $\cQMsq$ will only affect the CSS prediction of the presence of
disconnected hybrid star branches.

\section{Mass-radius relations}
\label{sec:mass-radius}

The CSS parameterization allows us to conveniently survey the
mass-radius relation ships for a large class of possible quark matter
equations of state. We will first discuss how the form of the
mass-radius relationship is determined by the CSS parameters, then we
will see where in the CSS parameter space the NJL quark matters EoSs
are to be found.

The general dependence of the mass-radius relationship on the CSS
parameters is shown schematically in Fig.~\ref{fig:schem}.  We have
fixed $\cQMsq$ and we vary $\ptrans/\etrans$ and $\De
\ep/\etrans$. There are four regions in the space of possible quark
matter EoSs corresponding to four topologies of the mass-radius curve
for compact stars: the hybrid branch may be connected to the nuclear
branch (C), or disconnected (D), or both connected and disconnected
branches may be present (B), or neither (A) \cite{Alford:2013aca}.
The mass-radius curve in each region is depicted in inset graphs, in
which the thick green line is the hadronic branch, the thin solid red
lines the stable hybrid stars, and the thin dashed red lines the
unstable hybrid stars.  The thick straight (red) line in
Fig.~\ref{fig:schem} is the critical line given by
Eq.~(\ref{eq:stabalf}), above which there is no connected hybrid
branch. We emphasize that this critical line is independent of the
speed of sound in the quark matter, and depends only on
$\De\ep/\etrans$ and $\etrans/\ptrans$.

\begin{figure*}[htb]
\parbox{0.5\hsize}{
\includegraphics[width=\hsize]{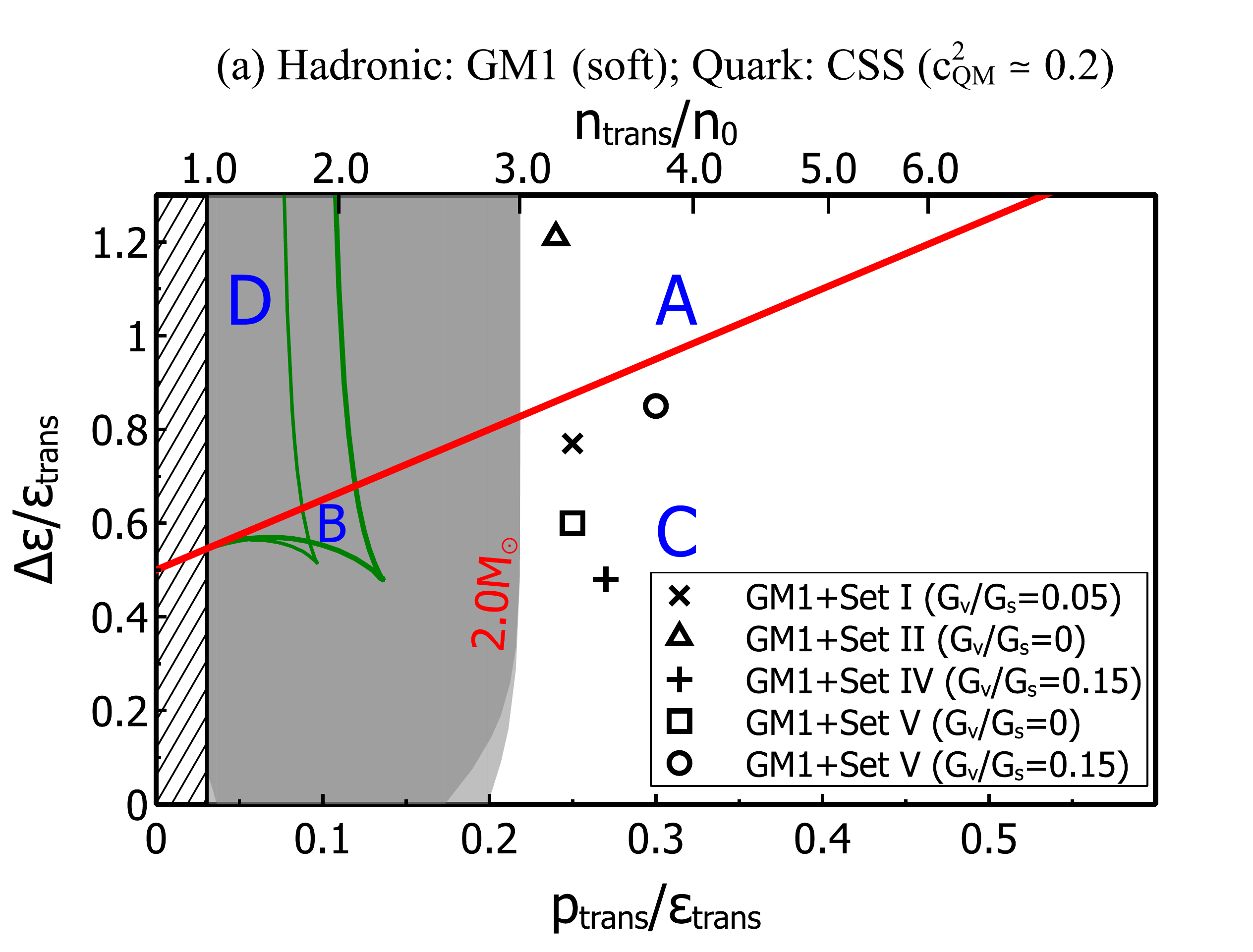}\\[0.5ex]
}\parbox{0.5\hsize}{
\includegraphics[width=\hsize]{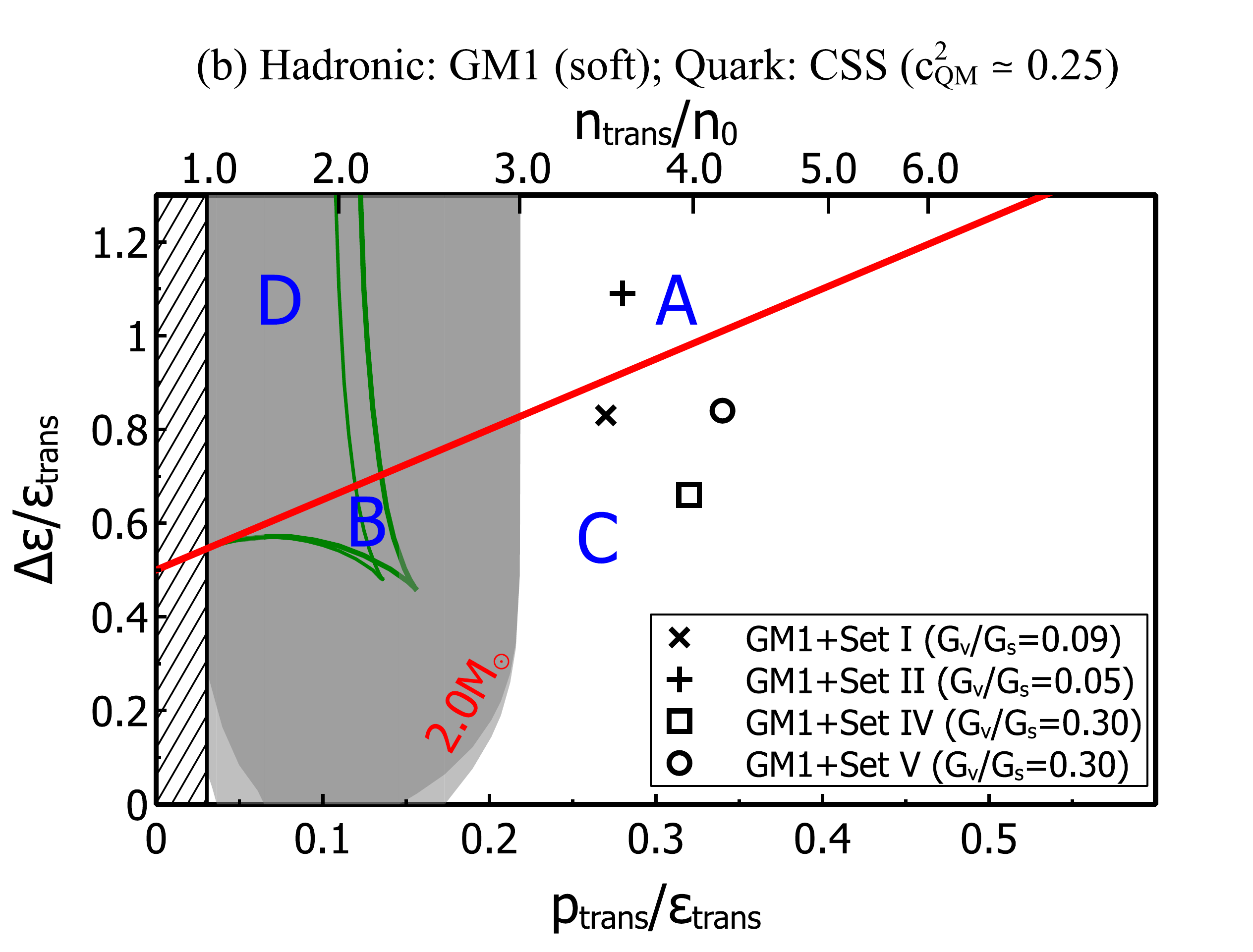}\\[0.5ex]
}\\[0.5ex]
\parbox{0.5\hsize}{
\includegraphics[width=\hsize]{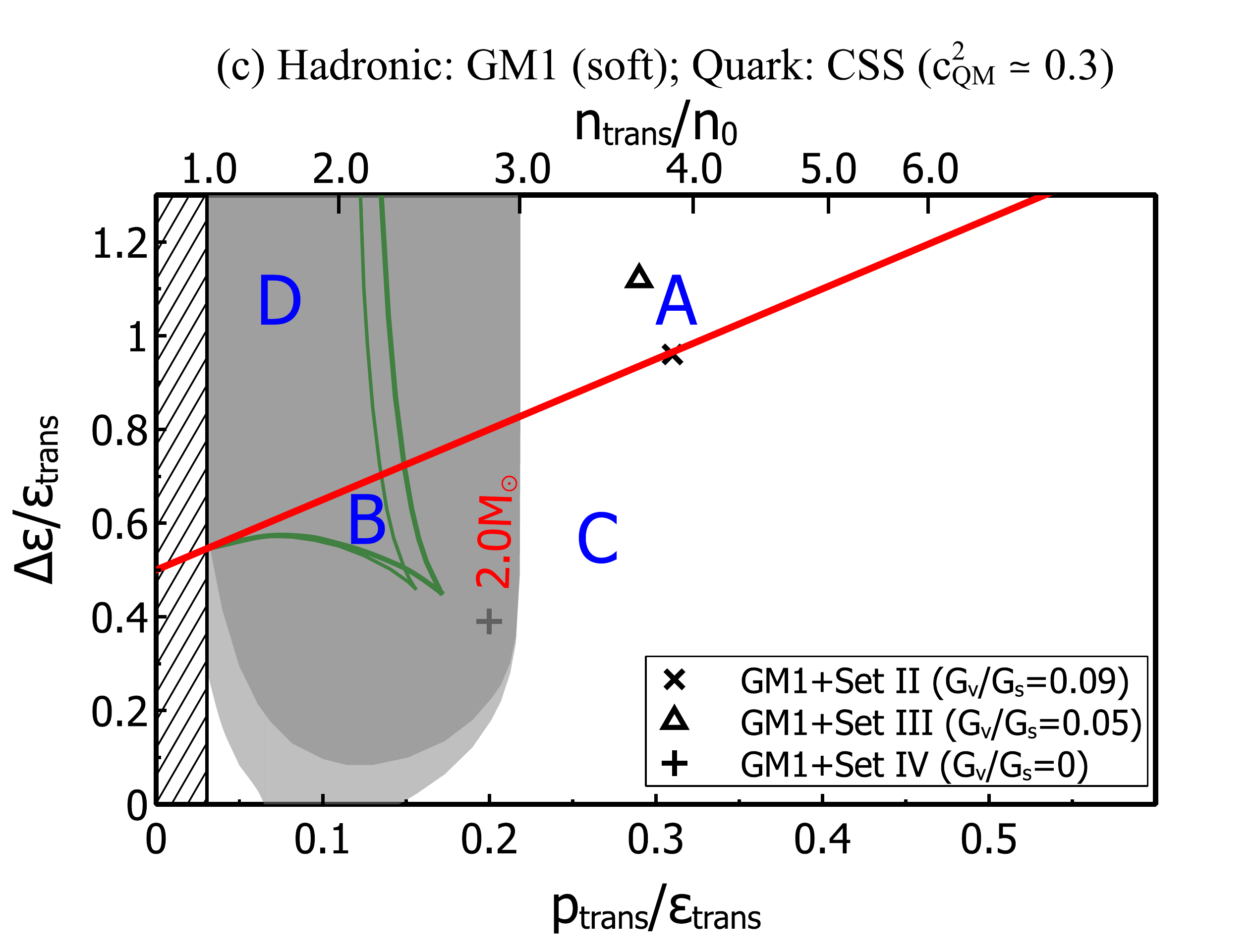}\\[0.5ex]
}\parbox{0.5\hsize}{
\includegraphics[width=\hsize]{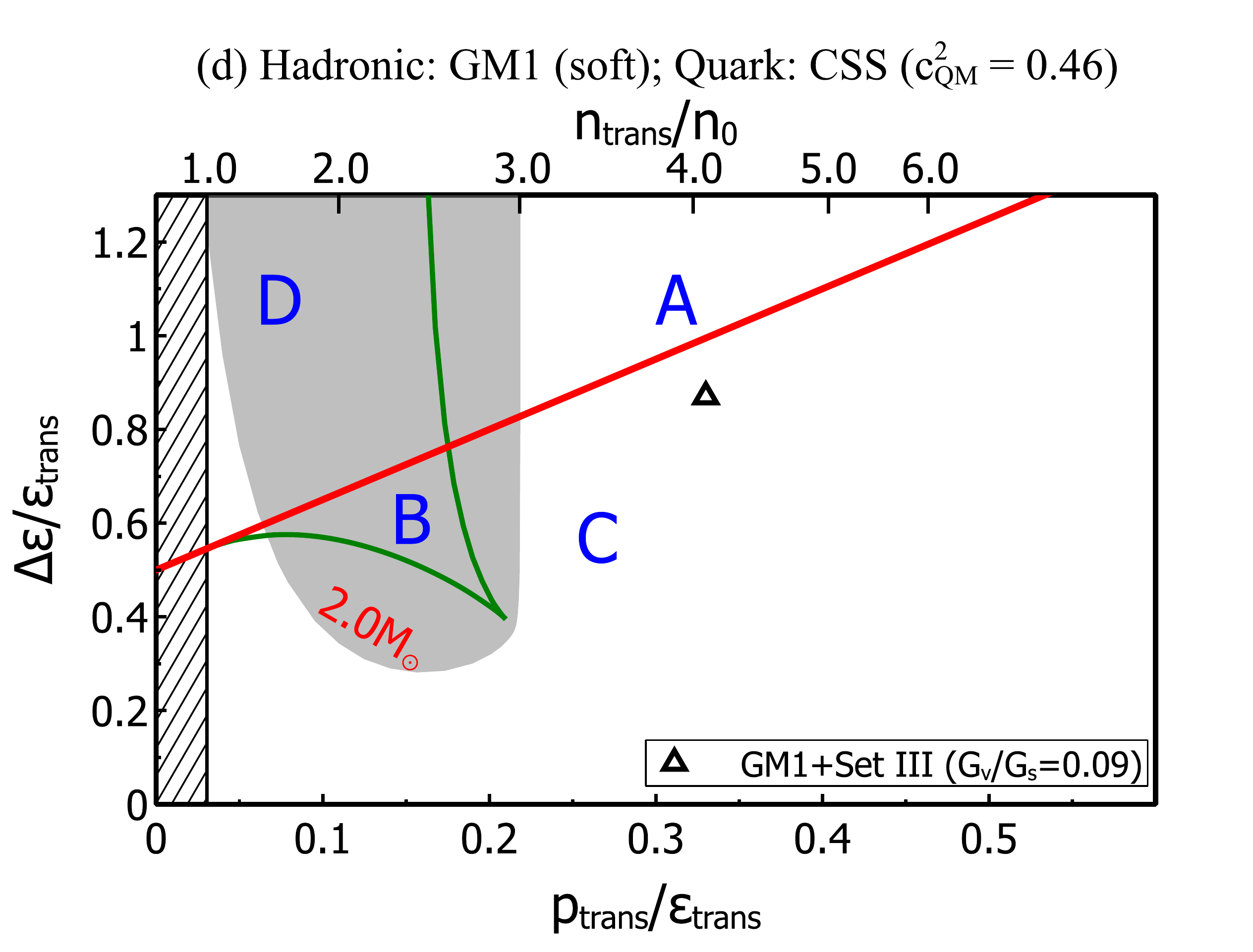}\\[0.5ex]
}
\caption{(Color online) Diagram showing (black symbols) where different
  quark matter parameterizations, for the local and nonlocal models, combined with GM1 nuclear matter, fall in the
  CSS parameter space. Each panel is for a different range of
  $\cQMsq$. EoSs below the straight solid (red) line (regions B and
  C) yield a connected hybrid branch. EoSs within the shaded gray
  area are excluded because their heaviest star is below
  $2\,\Msolar$. The hatched area at densities $n_{\rm trans} < n_0$
  is excluded because uniform nuclear matter is not
      stable in that region. See Tables \ref{table:nolocal} and
  \ref{table:local}.
\label{map-GM1}
}
\end{figure*}

\begin{figure*}[htb]
\parbox{0.5\hsize}{
\includegraphics[width=\hsize]{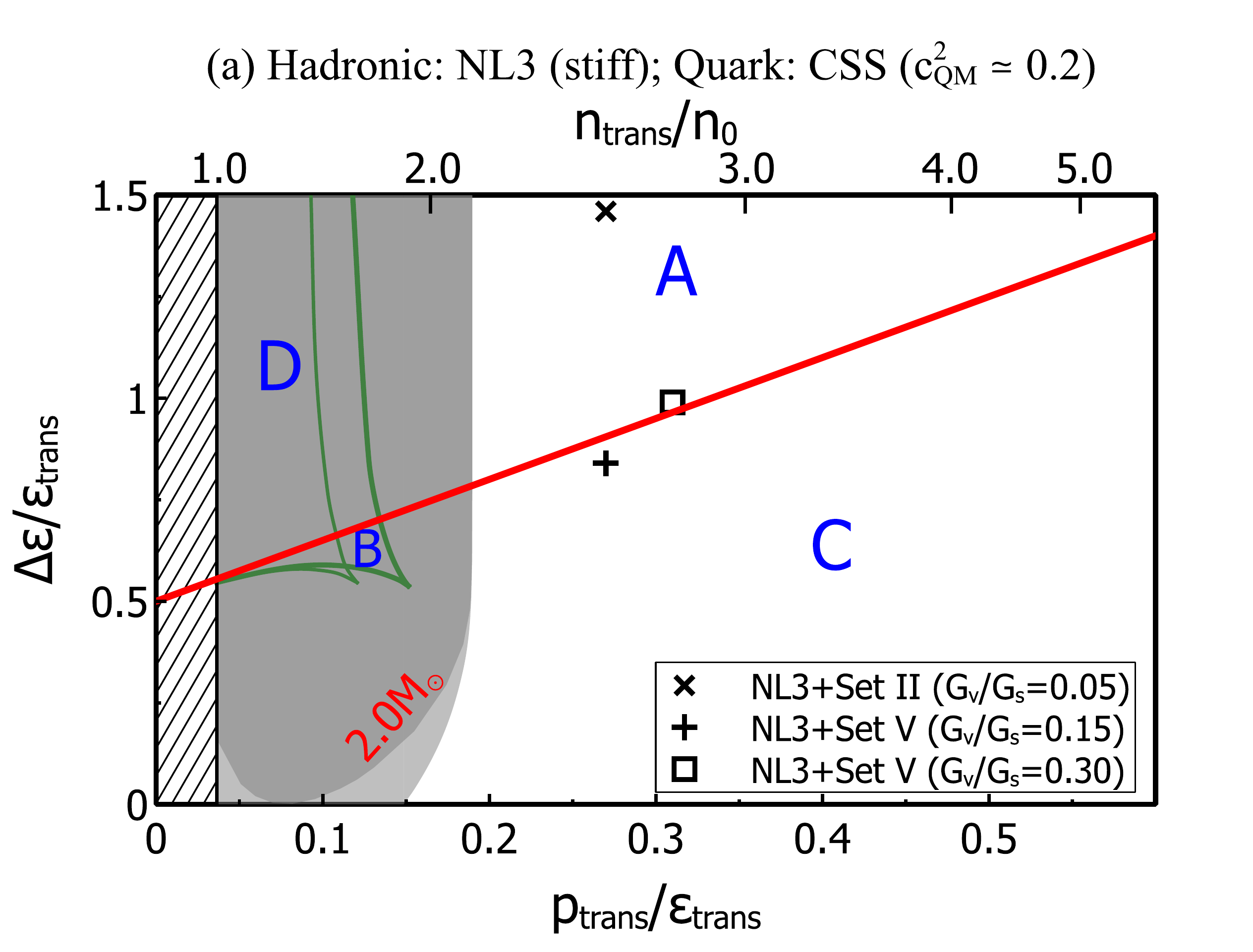}\\[0.5ex]
}\parbox{0.5\hsize}{
\includegraphics[width=\hsize]{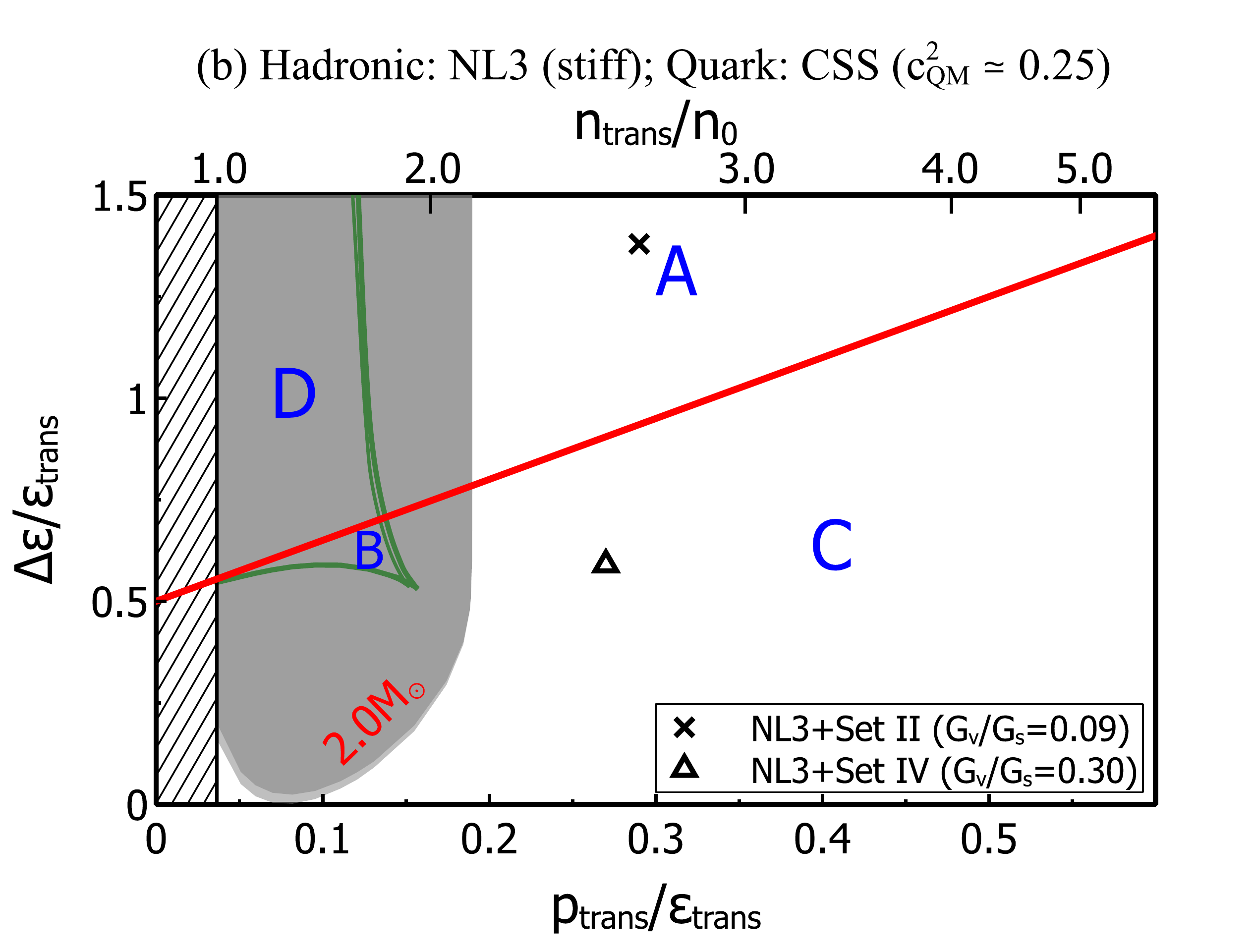}\\[0.5ex]
}\\[0.5ex]
\parbox{0.5\hsize}{
\includegraphics[width=\hsize]{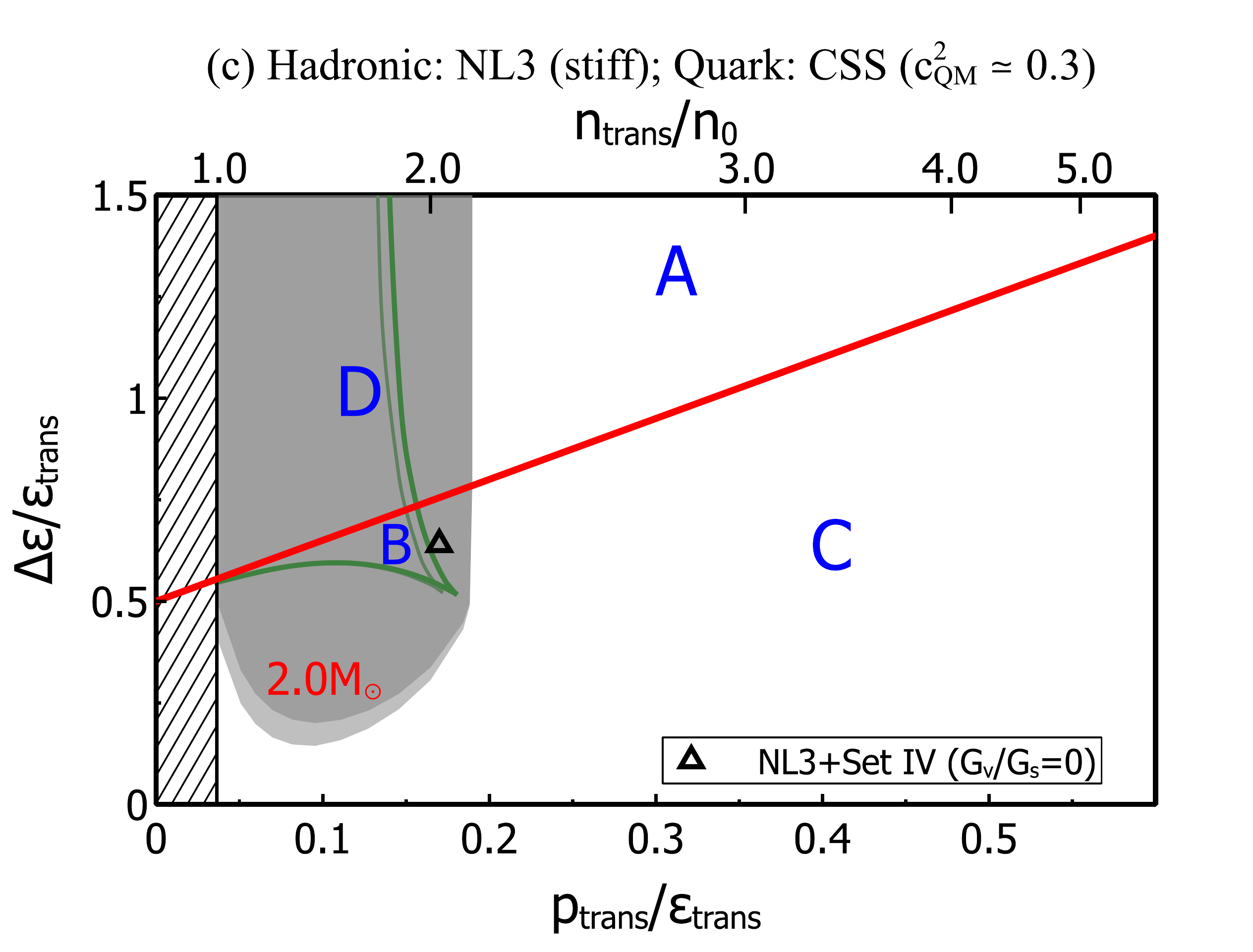}\\[0.5ex]
}\parbox{0.5\hsize}{
\includegraphics[width=\hsize]{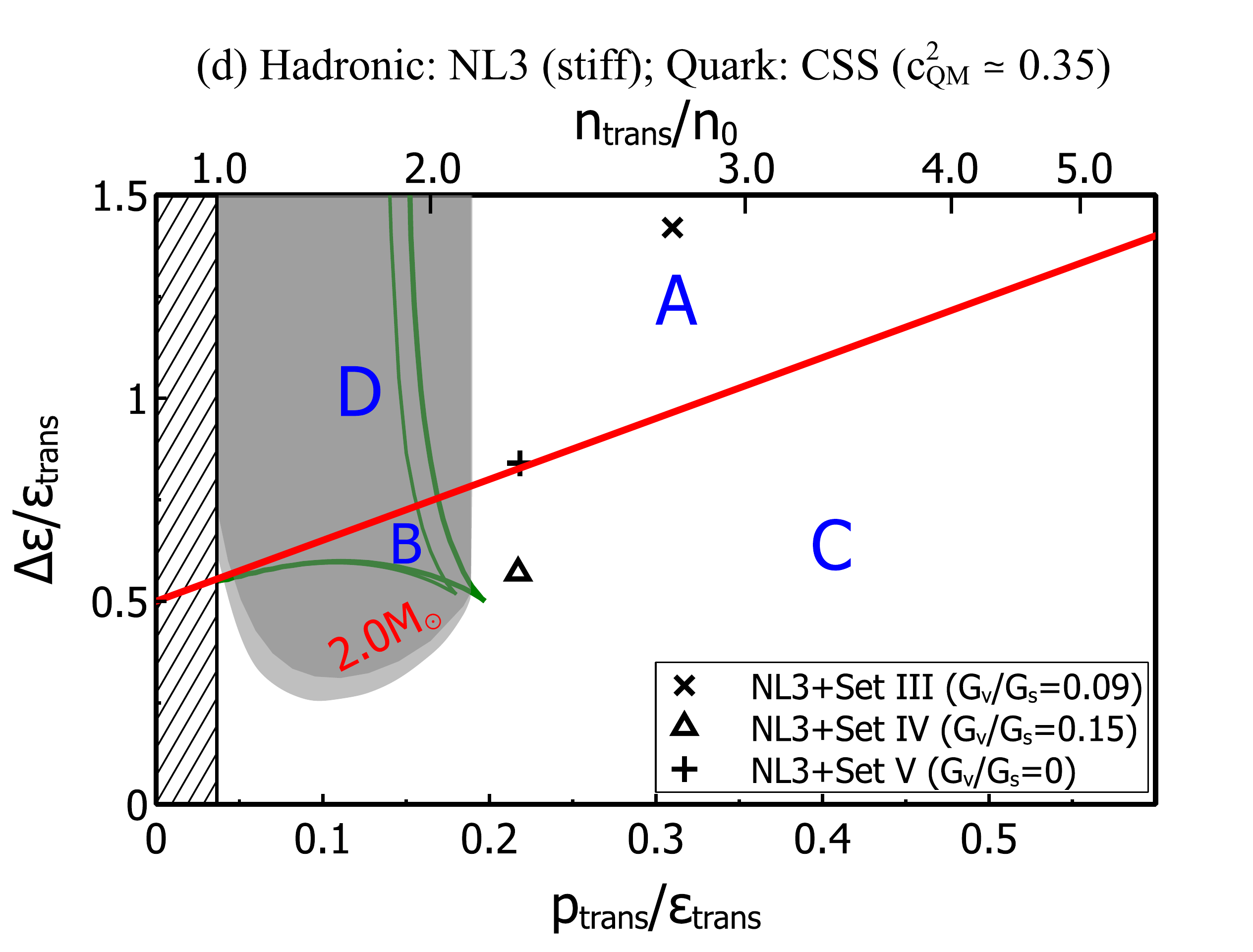}\\[0.5ex]
}
\caption{(Color online) Diagram showing (black symbols) where different
  quark matter parameterizations, for the local and nonlocal models, combined with GM1 nuclear matter, fall in the
  CSS parameter space. Each panel is for a different range of
  $\cQMsq$. EoSs below the straight solid (red) line (regions B and
  C) yield a connected hybrid branch. EoSs within the shaded gray
  area are excluded because their heaviest star is below $2\,\Msolar$.
  The hatched area at densities $n_{\rm trans} < n_0$ is
      excluded because uniform nuclear matter is not stable in that
      region. See Tables \ref{table:nolocal} and \ref{table:local}.
\label{map-NL3}
}
\end{figure*}


We see in Tables \ref{table:nolocal} and \ref{table:local} that each
NJL model of the quark matter EoS corresponds to a set of CSS
parameters, i.e. to a value of $\cQMsq$ and a point in the
accompanying $(\ptrans/\etrans, \De \ep/\etrans)$ plane.  In
Figs.~\ref{map-GM1} and \ref{map-NL3}, we use black symbols to plot
the CSS parameter values corresponding to the NJL models listed in
Tables \ref{table:nolocal} and \ref{table:local}.  In
Fig.~\ref{map-GM1} the NJL models for quark matter are combined with
nuclear matter described by the GM1 EoS. In Fig.~\ref{map-NL3} the
nuclear matter is described by the NL3 EoS.
To capture the variation in $\cQMsq$, each figure has 4 panels, each
one devoted to a small range of values of $\cQMsq$. For example, in
the panel (a)  of Fig.~\ref{map-GM1} we show the NJL models that
have $\cQMsq\approx 0.2$. One of them (the triangle-point
corresponding to GM1+Set I) is in the ``A'' region, meaning that it
gives no hybrid stars because its phase transition is so strongly
first-order.  The others are all in the ``C'' region, meaning that
their mass-radius relation includes a connected branch of hybrid
stars. However the connected branch is short, covering a
range of no more than $0.05\,\Msolar$.
(For other nuclear and quark models the hybrid branches are even shorter,
covering a range of order of $10^{-3}\,\Msolar$ or less.)
It would be difficult to detect such a short hybrid branch
in mass-radius observations.

In each panel in Figs.~\ref{map-GM1} and \ref{map-NL3}
there is a grey shaded region covering the equations of state
that are ruled out because they give a maximum compact star mass below
$2\,\Msolar$. The paler shading is for the lowest value of $\cQMsq$
that is associated with the panel, and the darker shading is for the
highest value of  $\cQMsq$.

The ``D'' region (where disconnected hybrid star
branches exist) and the ``B'' region (where both connected and
disconnected hybrid star branches exist) are inside the grey shaded
region, meaning that for the GM1 nuclear matter EoS and for quark
matter with $\cQMsq\approx 0.2$ all CSS-compatible quark matter EoSs
that would give disconnected hybrid branches are ruled out because
their heaviest star is too light.

The remaining panels of Fig.~\ref{map-NL3} show similar plots for the NJL
EoSs with higher sound speed values: $\cQMsq \approx 0.25,\,0.3$, and
finally one for the GM1+Set III EoS which has $\cQMsq=0.46$.  We see
that as $\cQMsq$ grows the B+D region (where disconnected hybrid
branches exist) also grows, but always remains within the grey
region, excluded by the $M_{\rm max}>2\,\Msolar$ constraint.

Most of the non-local NJL EoSs lack hybrid branches because the
energy density jump at the transition is so large that the quark
matter core immediately destabilizes the star. Even for the
quark EoSs that are in the C region below the red
line, the hybrid branch is very short. This is because those
EoSs have high transition pressure ($\ptrans/\etrans \gtrsim 0.2$)
so any star with central pressure
high enough to develop a quark matter core is already close to the
central pressure at which the hadronic star would become unstable
even without a quark matter core, so when the core appears it easily
``pushes the star over the edge'' into instability.

To evaluate the sensitivity of our results to other parameters
in the NJL model, we looked at two alternate parameter sets to see
if they yielded longer hybrid branches.

Firstly, to probe the effects of flavor mixing
we set the 'tHooft coupling constant $H$
in the IV parametrization to zero. This corresponds to a
local NJL model without flavor mixing.
To reproduce the vacuum constituent quark masses, we set
$G_S \Lambda^2 = 4.638$ and $m_s =112.0$\,MeV
\cite{Buballa:2003qv}. Since increasing $G_V$ tends to result in
smaller quark cores, we set $G_V = 0$. Even so, we found that
setting $H$ to zero further disfavors hybrid branches.
For GM1 hadronic matter the resultant
connected hybrid branches were even shorter
($\Delta M \sim 4 \times 10^{-4} \msun$)
than in the case with $H \neq 0$. For NL3 hadronic matter
there was no stable hybrid branch, where there had been a
very short one when $H \neq 0$.
We conclude that the 'tHooft flavor mixing term
favors the appearance of a hybrid stable
branch.

Secondly, to analyze how sensitive the results are to changes in the strange
quark mass we considered another parametrization for the local NJL model
without vector coupling, \cite{Klimt:1990ws} with $\Lambda = 750$\,MeV, $G_S^2
= 3.67$, $H \Lambda^5 = -8.54$, $m_u = m_d = 3.6$\,MeV and $m_s = 87$\,MeV.
With both GM1 and NL3 nuclear matter we found no hybrid branch.  This
could be attributed to the combination of two contributions: the lower
strange quark mass increases the fraction of strange quarks at lower
densities \cite{Buballa:2003qv} and a weaker flavor mixing favors a faster
increasing of strangness fraction with rising density. The combination of
these two contributions softens the quark matter EoS and helps to
destabilize hybrid stars. However, a color superconducting phase
in the core could help to stabilize the star \cite{Buballa:2003et}.

\section{Conclusions}
\label{sec:end}

We have studied the hybrid stars that arise from quark matter modeled
by local and non-local NJL models with vector interaction among the
quarks. The reason for studying the non-local NJL model is that it is
constructed to have the closest correspondence to QCD
\cite{Ripka:1997zb,GomezDumm:2006vz,Contrera:2007wu,Diakonov:1985eg,Schafer:1996wv,
  Roberts:1994dr,Roberts:2000aa,Parappilly:2005ei,Bowler:1994ir,Plant:1997jr,Broniowski:2001cx,Rezaeian:2004nf},
and has a natural cutoff in the form of the momentum dependence of the
effective quark masses.  For the equation of state of hadronic matter
we used the non-linear relativistic mean field model with two
different parameterizations, GM1 (softer) and NL3 (stiffer). We
assumed that the surface tension at the interface is high enough so
that there is a sharp interface between the phases with no mixed
phase.

The main physical conclusion is that the non-local NJL models that we
studied typically give no hybrid stars, while the local NJL models
sometimes give hybrid stars, but they cover a very small range of
masses and radii (this is different from the behavior seen when the
phase boundary is a mixed phase (Gibbs construction)
\cite{Orsaria:2012je,Orsaria:2013hna}).  According to Tables
\ref{table:nolocal} and \ref{table:local}, the mass range is of order
$10^{-3}\,\Msolar$ or less. One would expect a very small fraction of
observed neutron stars to be in the hybrid branch, and they would be
difficult to identify via mass and radius measurements, but it is
possible that those stars, which would have very small quark matter
cores, might nonetheless have distinctive observable properties. One
possibility is fast cooling if the quark matter core had a high
neutrino emissivity. Another is that the density discontinuity
associated with the quark core might affect the frequency spectrum of
non-radial oscillation modes \cite{Miniutti:2002bh}, leading to
signatures in the gravitational wave emission.

We found that in the NJL models that we studied, the EoSs for quark
matter are adequately represented by the CSS (Constant-Speed-of-Sound)
parameterization, and that the CSS parameterization helps us to
understand the characteristics of the hybrid stars predicted by the
NJL model. As we see in Figs.~\ref{map-GM1} and \ref{map-NL3}, and in
Fig.~2 of Ref.~\cite{Alford:2015dpa}, quark matter EoSs that are
soft, with $\cQMsq\lesssim 1/3$, are severely restricted by the
$2\,\Msolar$ constraint.  To meet it they must have a high transition
pressure.  When the transition pressure is high, approaching the
pressure at which the hadronic star would become unstable anyway even
without a transition to quark matter, the hybrid branch tends to be
very short because, unless the transition to quark matter is very
weakly first-order, the appearance of the denser core of quark matter
pushes the star close to the point of instability. This explains why
in this work we only find very short connected hybrid branches and no
disconnected hybrid branches.  The NJL models studied here tend to
have a low speed of sound.  The $2\,\Msolar$ constraint then rules out
disconnected branches (the B and D regions of Figs.~\ref{map-GM1} and
\ref{map-NL3}) and requires a high transition pressure.

The fact that the hybrid branches are so short means that only a very
small range of pressures above the transition to quark matter are
physically attained in hybrid stars, so, since the CSS
parameterization is essentially a Taylor expansion of $\ep(p)$ around
the transition, it is a very accurate representation of the quark
matter EoS over this range.

The CSS parameterization exposes the physical consequences of varying
the parameters of the NJL quark matter EoS. Tables \ref{table:nolocal}
and \ref{table:local} show the mapping from the NJL parameters to the
CSS parameters. We see that\\ [1ex] $\bullet$ Increasing $m_s$ at fixed
vector to scalar coupling ratio, which corresponds to going from Set I
to III (non-local NJL) or IV to V (local NJL), leads to an EoS with
higher phase transition pressure, a larger energy density
discontinuity at the transition, and higher speed of sound in the
quark matter.\\ [1ex] $\bullet$ Increasing the vector to scalar coupling
ratio with other parameters held fixed also increases the transition
pressure, energy density discontinuity, and speed of
sound.\\ [1ex] $\bullet$ The consequences for hybrid stars can be read off
from Figs.~\ref{map-GM1} and \ref{map-NL3}. Increasing the strange
quark mass or the repulsive vector interaction does not favor hybrid
stars since at higher transition pressure and energy density
discontinuity it becomes more likely that a quark matter core will
destabilize the star.

 In the future, it would be useful to study further variations on the
models considered here. This will help to shed light on how
generic our results are. For hadronic matter, it would be interesting
to use the DD2-EV EoS
\cite{Benic:2014jia} which is a relativistic model that has been
calibrated to fit known properties of nuclear matter.
For quark matter, one should study the role of the mixing ('tHooft)
term.
It is known that the 'tHooft
interaction shifts the critical end point location in the QCD phase
diagram, affecting the nature of the hadron-quark phase transition
(see for example Refs.~\cite{Klimt:1990ws} and
 ~\cite{Fukushima:2008wg}). In addition, one could include a diquark
coupling. The presence of diquark condensates lowers
the transition chemical potential, allowing a lower transition
pressure, which has been seen to yield longer hybrid branches than we
found in this work \cite{Klahn:2006iw}.

\section{Acknowledgments}

FW is supported by the U.S. National Science Foundation under grant
no.\ PHY-1411708.  GAC and MGO acknowledge financial support by
CONICET and UNLP, Argentina. IFR-S is a fellow of CONICET. MGA and SH
are supported by the U.S.\ Department of Energy, Office of Science,
Office of Nuclear Physics under Award Number \#DE-FG02-05ER41375, and
by the DOE Topical Collaboration ``Neutrinos and Nucleosynthesis in
Hot and Dense Matter'' contract \#DE-SC0004955.

\renewcommand{\href}[2]{#2}

\newcommand{\apjl}{Astrophys. J. Lett.\ }
\newcommand{\mnras}{Mon. Not. R. Astron. Soc.\ }
\newcommand{\aap}{Astron. Astrophys.\ }

\bibliographystyle{JHEP_MGA}
\bibliography{HS}   
\end{document}